\def\lesssim{{_ <\atop{^\sim}}}

\def\ap3m{AP$^3$M}
\def\LCDM{$\Lambda$CDM}

\def\hMsun{$h^{-1}{\ }{\rm M_{\odot}}$}

\def\nbody{$N$-body}
\def\c15{$c_{\rm 1/5}$}

\newcommand{\Eq}[1]{equation~(\ref{#1})}
\newcommand{\Fig}[1]{Fig.~\ref{#1}}
\newcommand{\mlapm}{\texttt{MLAPM}}

\def\ea{et~al.~}                            

\def\lesssim{\mathrel{\hbox{\rlap{\hbox{\lower4pt\hbox{$\sim$}}}\hbox{$<$}}}}
\def\gtrsim{\mathrel{\hbox{\rlap{\hbox{\lower4pt\hbox{$\sim$}}}\hbox{$>$}}}}

\documentclass[useAMS,usenatbib]{mn2e}
\include{epsf}

\title[Mapping Substructures in Dark Matter Halos]
      {Mapping Substructures in Dark Matter Halos}

\author[Knebe~\ea]
       {Alexander Knebe$^{1,2}$, 
        Stuart P.~D. Gill$^2$, Daisuke Kawata$^2$, Brad K. Gibson$^2$ \\
        $^1$Astrophysikalisches Institut Potsdam, 
            An der Sternwarte 16, 14482 Potsdam, Germany\\
        $^2$Centre for Astrophysics~\& Supercomputing, 
            Swinburne University, Mail \#31, P.O. Box 218, 
            Hawthorn, Victoria, 3122, Australia
       }
\begin{document}

\date{Received ...; accepted ...}
\pagerange{\pageref{firstpage}--\pageref{lastpage}} \pubyear{2004}

\maketitle

\label{firstpage}

\begin{abstract}

We present a detailed study of the real and integrals-of-motion space
distributions of a disrupting satellite obtained from a fully
self-consistent high-resolution cosmological simulation of a galaxy
cluster. The satellite has been re-simulated using various analytical
halo potentials and we find that its debris appears as a coherent
structure in integrals-of-motion space in all models (``live'' and
analytical potential) although the distribution is significantly
smeared for the live host halo.  The primary mechanism for the
dispersion is the mass growth of the host.  However, when
quantitatively comparing the effects of ``live'' and time-varying host
potentials we conclude that not all of the dispersion can be accounted
for by the steady growth of the host's mass. We ascribe the remaining
discrepancies to additional effects in the ``live'' halo such as
non-sphericity of the host and interactions with other satellites
which have not been modeled analytically.

\end{abstract}

\begin{keywords}
galaxies: clusters -- galaxies: formation -- galaxies: evolution -- 
n-body simulations
\end{keywords}

\section{Introduction}

With the discovery of stellar streams within the Milky Way 
\citep[e.g][]{HWdZ99,CB00,ILIC02,BKGF03,YNG03,NHF04,MKL04} and 
M31 \citep[e.g.][]{IILFT01,MIL04}
and streams and shells in clusters 
\citep[e.g.][]{TM98,GW98,CMB00,FMMRH02} they
have become a standard fixture in our understanding of galaxy and
cluster formation.  These streams provide important observational
support for the hierarchical build-up of galaxies and clusters and the
$\Lambda$-dominated cold dark matter ($\Lambda$CDM) paradigm. Future
observational experiments such as RAVE \footnote{\tt
http://astronomy.swin.edu.au/RAVE/} and GAIA\footnote{\tt
http://astro.estec.esa.nl/GAIA/} are designed to shed further light on
the importance of streams in galaxy formation. Thus to better
interpret this observational data coming online in the near future it
is necessary to have clear theoretical understanding of the stellar
streams left behind by dissolving satellites.

The most common approach to this has been to simulate the disruption
of individual satellite galaxies in {\it static} analytical potentials
representative of the distribution of dark matter (DM) halos
\citep[e.g.][]{KN97,HdZ00,HMO01,ILIC02,BCDS03,MKL04}.
Such simulations have provided great insight into the
formation of stellar \citep[e.g.][]{ILIC02} 
and gaseous \citep[e.g.][]{YN03,CKMG04} streams, 
and helped to constrain the
shape of the Milky Way's halo \citep[e.g.][]{JZSH99,ILITQ01} 
and substructure content \citep[e.g.][]{JSH02}.

The \LCDM\ structure formation scenario predicts that small objects
form first and subsequently merge to form entities thus 
DM halos are never ``at rest'' but always be in the process of
accreting material from its vicinity and hence grow in mass. This also
extends to the Milky Way, although it encountered its last
major merger some 10 Gyrs ago
\citep[e.g.][]{GWN03}.  Therefore, the question presents itself, 
{\it is a static analytical DM potential a valid assumption for a study
of the disruption of satellites?}

While Zhao~\ea (1999) already addressed the issue of the evolution of
a satellite galaxy in a time-varying \textit{analytical} potential, we
are going to complement this study by comparing the
disruption processes of satellite galaxies in analytical potentials
\textit{and} fully self-consistent cosmological simulations.

\section{The Simulations}\label{simulations}


This study focuses on halo~\#1 from the series of high-resolution
\nbody\ galaxy clusters simulations \citep{GKG04a,GKGD04}.
Halo~\#1 was chosen because it is the oldest of the simulated clusters
(i.e. 8.3 Gyrs) having a reasonably quiet merger history.  The
self-consistent cosmological simulations were carried out using the
publicly available adaptive mesh refinement code
\mlapm\ \citep{KGB01} in a standard \LCDM\ cosmology ($\Omega_0 =
0.3,\Omega_\lambda = 0.7,
\Omega_b h^2 = 0.04, h = 0.7, \sigma_8 = 0.9$). 
For a more details and an elaborate study of these simulations we
refer to reader to \citet{GKG04a,GKGD04}.

From this simulation we chose one particular satellite galaxy orbiting
within the live host halo and refer to its evolution as the ``live
model''.  This choice was based upon two constraints: firstly, the
satellite contains a sufficient number of particles and secondly, it
has had multiple orbits.  The satellite used throughout this study
complies with these criteria in a way that it consists of $\sim$
15,000 particles (as opposed to $\sim$ 800,000 for the host) at the
``initial'' redshift $z=1.16$ (8.3 Gyrs ago) and has roughly 4 orbits
within the host's virial radius until $z=0$ when the mass of the host
reached $2.8\times 10^{14}$\hMsun\ (roughly 1,550,000 particles).

\begin{figure}
\begin{center}
\epsfxsize=5cm
\epsfbox{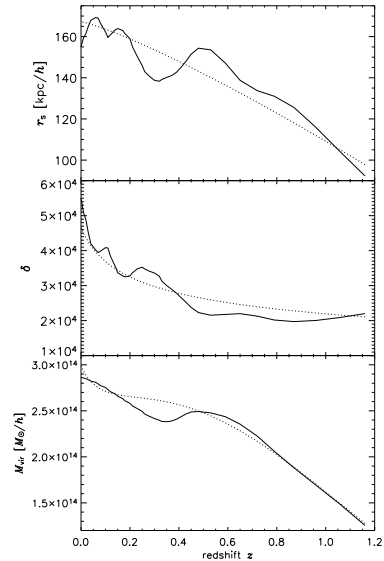}
\caption{The redshift dependence for $r_s$ (upper panel),
 $\delta_s$ (middle panel) and consequentially 
 virial mass, $M_{\rm vir}$ (lower panel). 
 The solid lines indicate the evolution
 in the live model, and the dashed lines are our best fitting formula
 to describe the evolution (see eq.\ \ref{HaloFit}).}
\label{HostModel}
\end{center}
\end{figure}

After extracting the satellite from the cosmological simulation we
used a tree $N$-body code \citep[GCD+:][]{KG03} to model its evolution
for 8.3 Gyrs in two (external) analytical potentials. The first of
these two models we call the ``evolutionary model'' which uses an
analytical reconstruction of the live halo's potential as follows: at
each available snapshot of the live model, we fitted the host's DM
density profile to the functional form of a (spherical) Navarro,
Frenk~\& White profile
\citep[][NFW]{NFW97}:
\begin{equation}
 \frac{\rho(r)}{\rho_b} = \frac{\delta_s}{(r/r_s) (1+r/r_s)^2} \ ,
\end{equation}
where $\rho_b$ measures the cosmological background density,
$\delta_s$ controls the amplitude and $r_s$ measures the radius where
the profile turns from its logarithmic slope of $d\log\rho/d\log r =
-1$ to $d\log\rho/d\log r = -3$.  From this series of snapshot NFW
profiles we reconstruct the evolution of the parameters $\delta_s (z)$
and $r_s (z)$. Our best fitting functions are given below:
\begin{equation} \label{HaloFit}
 \begin{array}{lcl}
  \delta_s & = & \displaystyle \frac{23456}{(z+0.08)^{0.28}} - 1015 \ ,\\
\\
   r_s     & = & 167.28 - 58.14 \ z^{1.2} \ .\\
 \end{array}
\end{equation}
\Fig{HostModel} shows the evolution of $r_s$,
$\delta_s$ and consequentially mass as a
function of redshift for both the numerical simulation
alongside the analytical formula described by \Eq{HaloFit}.
The increase of the host's mass around $z=0.45$ is related to
a transient (``backsplash'') satellite galaxy passing just within the
virial radius at high velocity \citep[cf.][]{GKG04b}.
The second model assumes a static analytical host potential and hence
is labeled ``fixed model''.  The parameters $\delta_s= 21070$ and
$r_s=97.8$ adopted for the fixed model agree with the initial values
for the evolutionary model.
Note that in both analytical models the hosts are assumed to be
spherical. Dynamical friction is not implemented, either.

\section{The Results}\label{Analysis}

\begin{figure}
\begin{center}
\epsfxsize=7.5cm
\epsfbox{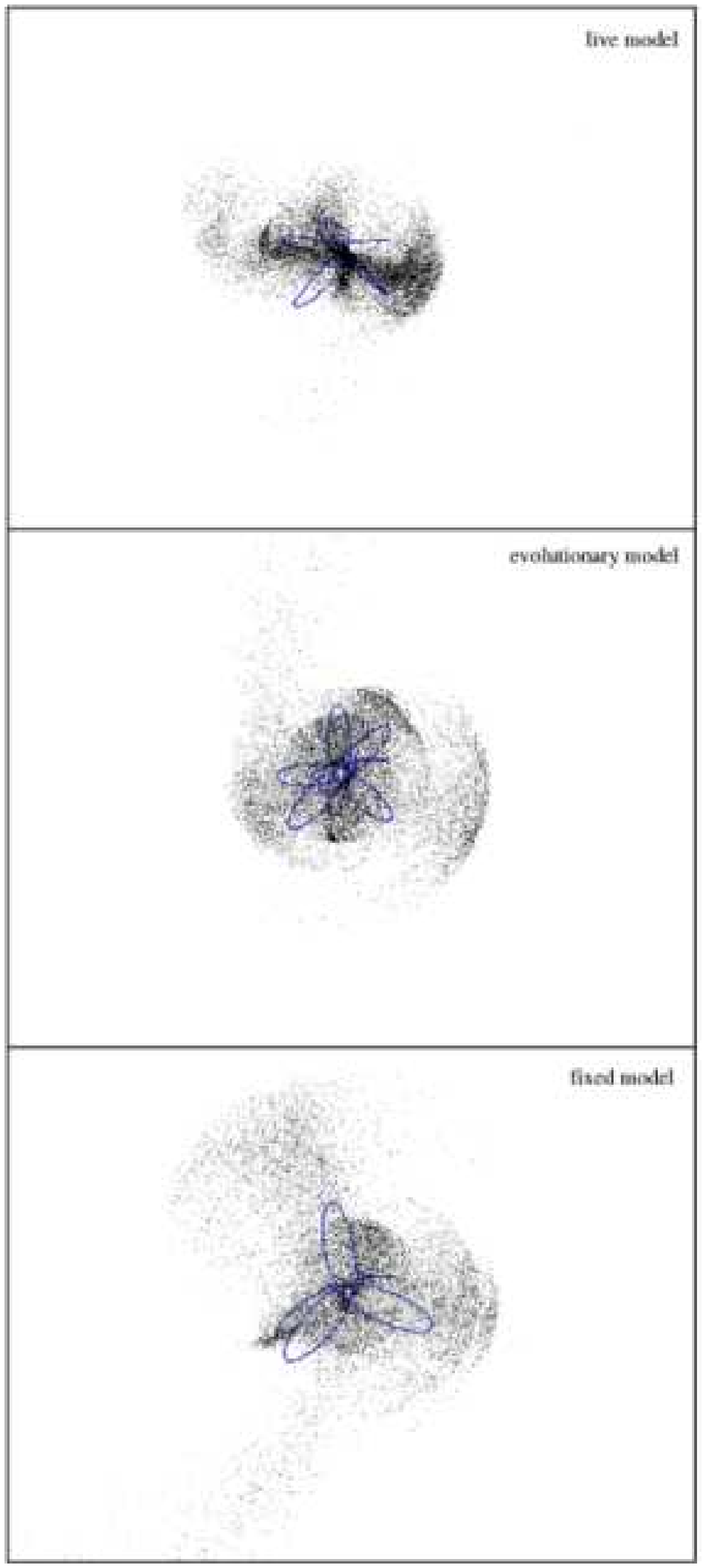}
 \caption{ The projected particle distribution for all the models
           at $z=0$. The lines indicate the orbital paths of the satellite
           of the respective model.}
\label{SatNow}
\end{center}
\end{figure}

In \Fig{SatNow} we show the real-space distribution (left panel) of
the disrupted satellite after the 8.3 Gyrs evolution, i.e.\ at $z=0$
along with its distance to the host as a function of time (right
panel).  The spread in the number of orbits amongst the models can
readily ascribed to the difference in host mass and the small
``mis-modeling'' of its growth as seen in \Fig{HostModel} at around
redshift $z=0.3$.

This figure further highlights a number of interesting differences and
similarities amongst the models.  The most striking feature is that
neither of the analytical models is capable of producing the
real-space distribution of particles seen in the live model, and the
live model shows the most ``compact'' distribution.  Although all
models display ``shell''-like features, they are less eminent for the
live host. Instead, the live model exhibits a ``cross-like'' feature
and appears to be more compact in the central region, respectively. A
comparable feature, in fact, has also been noted in observational
streams
\citep[cf.][]{HGKG04}.

\citet{HdZ00}
outlined a method for identifying stellar streams within observational
data sets by utilising conservation of energy, $E$, and angular
momentum, $L$, (i.e. the integrals-of-motion) for spherically symmetric
and time-independent potentials. In fact, this method proved to be a
powerful tool for identifying streams in the Milky Way using the proper
motions of solar neighbour stars
\citep[e.g.][]{HWdZ99,CB00,BKGF03,NHF04}.  We expand their analysis not
only limiting ourselves to static analytical models but extending the
study to include both our live and evolutionary models.

\begin{figure*}
\epsfxsize=15.2cm
\epsfbox{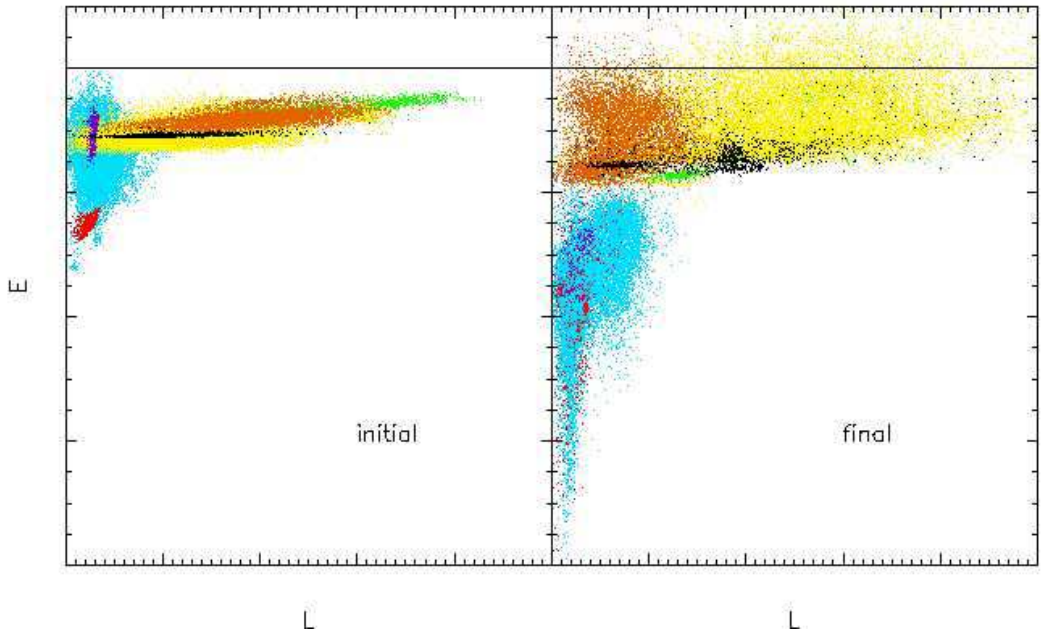}
\caption{
 The distribution of satellite particles in the $E-L$ plane
for the live models. The left panel shows the distributions
at the time the respective satellite galaxy enters the virial 
radius of the host whereas the right panel presents the 
distributions at $z=0$.
Different colours represent particles of different satellites.
}
\label{liveEL}
\end{figure*}

Before exploring the evolution in the $E-L$ plane (hereafter also
called ``integral-space'') of our target satellite in all three
models, we present the evolution for eight different satellites in the
live model alone. The result can be viewed in \Fig{liveEL} where each
satellite is represented by an individual colour with our target
satellite plotted as cyan.  The satellites shown in \Fig{liveEL} are
all taken from the self-consistent cosmological simulation and
represent a fair sample of all 158 available satellites, i.e. their
masses cover a range from from $M\sim 5\times 10^{10}$\hMsun\ to
$M\sim 4\times 10^{12}$\hMsun\ and they are on different orbits.  The
left panel of \Fig{liveEL} shows the distributions at the time the
satellite enters the virial radius of the host, whereas the right
panel displays the distributions at $z=0$.

\Fig{liveEL} allows us to gauge the ``drift'' of  satellites in
integral-space.  We note that, compared to Fig.\ 4 of
\citet{HdZ00}, the integrals-of-motion are hardly conserved in the
live model, neither for high or low mass streamers. The distributions
rather show a large scatter, and have been significantly ``re-shaped''
over time. In addition, the mean values of $E$ and $L$ are also moved
after the evolution.  For instance, the ``red'' satellite drifts in
time over to the initial position of the ``cyan'' satellite.
\Fig{liveEL} further demonstrates that the drift of our target
satellite (cyan dots) is comparable to the evolution of the other
satellites, thus indicating that this target satellite is a ``typical
satellite'' in that respect.

One encouraging result implied from \Fig{liveEL}, however, is that
even though the integrals-of-motion are changing over time,
\textit{satellites still appear coherent in the $E-L$ plane}.  Hence,
the integral-space analysis pioneered by \citet{HdZ00} still proves to
be a useful diagnostic to identify streams.  We further like to stress
that observations only provide us with the snapshot of the
distribution at today's time, i.e.\ the right hand panel of
\Fig{liveEL} and hence measuring ``evolution'' is beyond the scope of
RAVE and GAIA.

\begin{figure}
\epsfxsize=\hsize
\epsfbox{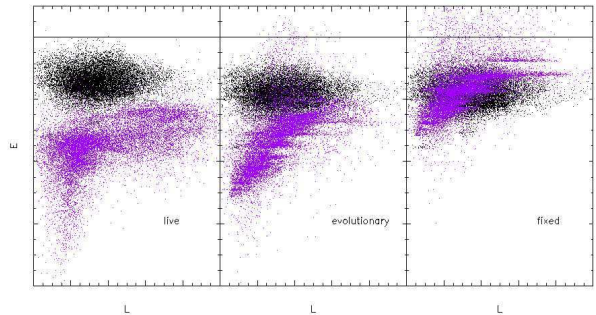}
 \caption{Total energy, $E$, versus absolute value of the angular
 momentum, $L$, for all satellite particles within
 the target satellite at initial redshift $z=1.16$ (black)
 and at $z=0$ (blue) in the live (left), evolutionary
 (middle) and fixed (right) models. The horizontal line indicates $E=0$.
}
\label{EL}
\end{figure}

\Fig{EL} now focuses on our target satellite alone, showing
its particles at initial and final time in the $E-L$ plane for all
three models, i.e.\ the live, evolutionary and fixed model.  We notice that
particles tend to form ``stripes'' in the $E-L$ plane (parallel to the
$L$-axis) indicative of a spread in angular momentum for particles of
comparable energy. These stripes form over time and have been directly
linked to both apo- and peri-centre passages of the satellite. The live
model deviates most prominently from the other models not only
reducing total energies, but also ``randomising'' the angular momentum
and hence lacking the prominence of these stripes.  We also observe in
\Fig{EL} that the fixed model is the only model to show a noticeable number 
of particles ($\approx$ 3\%) \textit{not} bound to the host halo,
i.e.\ $E>0$. This can be ascribed to the lower mass of the host halo.

\begin{figure}
\epsfxsize=\hsize
\epsfbox{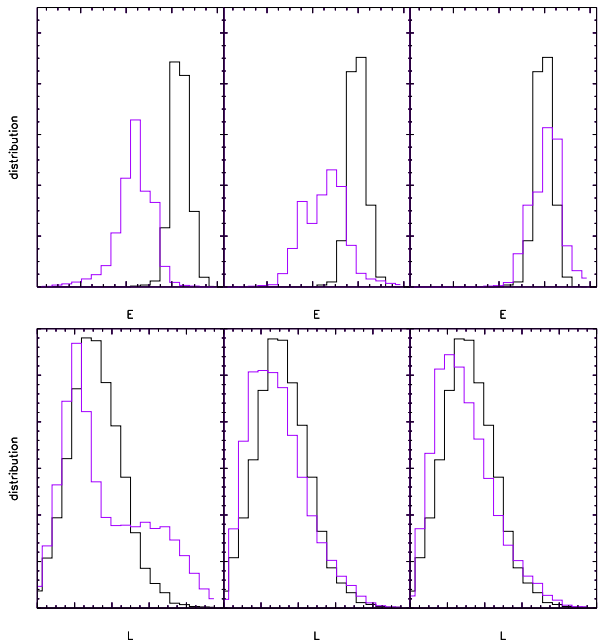}
 \caption{The distribution of energies (upper panels) at initial
          (black) and final (blue) time and angular momentum
          (lower panels) for the three models
          (live, evolutionary, fixed, from left to right).
} 
\label{ELdistrib}
\end{figure}

Moreover, \Fig{EL} confirms the findings of \citet{HdZ00} that
within a fixed host potential a satellite galaxies retains its
identity in the $E-L$ plane and the integrals-of-motion hardly
change, respectively.  This statement is further strengthened by
\Fig{ELdistrib} which presents the actual frequency distribution of 
$E$ and $L$ at the initial and final time.  

Not surprisingly, the live
model shows the largest deviations from the initial configuration 
(Fig.\ \ref{ELdistrib}). 
In the live model the distribution of $E$ and $L$ is broadened over time, and
the peak of $E$ systematically moves toward lower values, while the
peak of $L$ does not change dramatically for this particular
satellite.  This ``drop'' of $E$ is also observed in the evolutionary
model; it reflects the steady mass growth and related deepening of the
host potential (cf. \Fig{HostModel}).  Since the drift in $E$ is the
most significant change, the evolutionary model displays a similar
distribution to the live model in the $E-L$ plane at $z=0$.

\begin{figure}
\epsfxsize=\hsize
\epsfbox{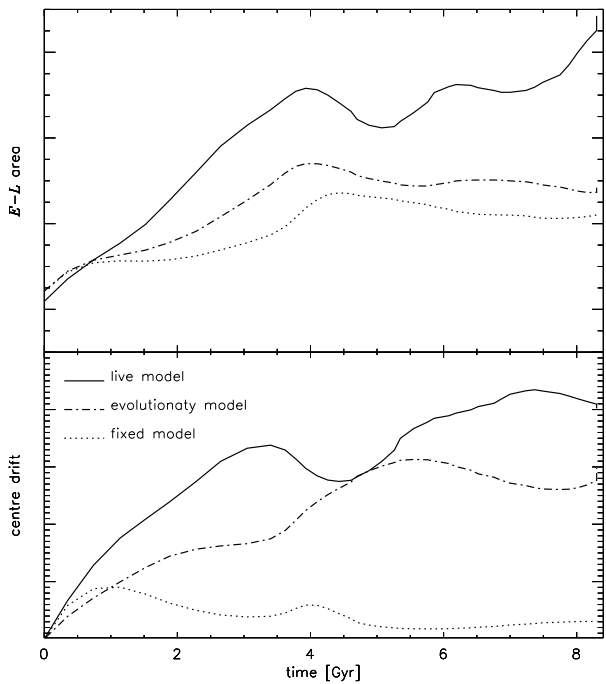}
 \caption{ The evolution of the area (top panel) and
the centre (bottom panel) in integral-space (see text for more details)
as a function of time for the live (solid), evolutionary
(dot-dashed) and fixed (dotted line) models.} 
\label{ELevolve}
\end{figure}

To further quantify the evolution of the satellite in integral-space,
we calculate the centre and area of the particle distribution seen in
\Fig{EL} as a function of time.  The centre is defined to be the two
dimensional arithmetic mean in $E$ and $L$ values, respectively. The
area covered in the $E-L$ plane is computed using a regular 128$^3$
grid covering that plane from $L_{\rm min}$ to $L_{\rm max}$ and
$E_{\rm min}$ to $E_{\rm max}$ in the respective model. Here, only
cells containing more than three particles are taken into account. The
evolution of both area and centre as a function of time since $z=1.16$
is presented in \Fig{ELevolve}.

The top panel of \Fig{ELevolve} (showing the change in area)
demonstrates that all three halos loose coherence of the distribution
in integral-space. This loss is most prominent for the live model and
least for the fixed model. This phenomenon can also be seen in a
comparison between Fig.\ 4 of
\citet{HdZ00} and the right panel of \Fig{liveEL}.
Therefore, we conclude that a fixed (and even an evolutionary) model
significantly underestimates the scatter in the integral-space, and
earlier predictions based on the simulation using a static (or
time-varying) analytical description for the halo potential
overestimate their efficiency of the detection ability of streams.

The bottom panel of \Fig{ELevolve} (showing the drift of the
distribution's centre in the $E-L$ plane with respects to its initial
position) is yet another proof that the fixed model holds the best
conservation of energy and angular momentum.  The ``live'' nature of
the host mass can be held responsible for not only the increased
dispersion (i.e. area) but also for the actual drift in
integral-space. We further like to stress that at not time the
evolutionary model matches the effects of the ``live'' model. There
always appears to be a well pronounced discrepancy which can be
ascribed to effects not considered in the evolutionary model, such as
the triaxiality of the host and interactions of the debris with other
substructure.  This leads to the immediate conclusion that for an
understanding of the formation of streams in integrals-of-motion space
it appears rather crucial to take into account the full complex
evolutioary effects of the host. The first such step has been taken by
Zhao~\ea (1999) who modeled the evolution of tidal debris in a
time-varying potential. However, our study further indicates that in a
``live'' scenario there are additional effects at work leading to an
even greater dispersion.

\section{Conclusions} \label{Conclusions}

In hierarchical structure formation scenarios as favoured by recent
estimates of the cosmological parameters \citep{SVP03} DM halos
continuously grow via both merger activity and steady accretion of
material. Furthermore, they also contain a great deal of substructure
and are far from spherical symmetry. We have investigated the
development of streams due to the tidal disruption of satellite
galaxies in a DM halo forming in a fully self-consistent cosmological
simulation. This not only models the time-dependency of the potential
(reflecting the mass growth of the host), it also accounts for other
effects such as the triaxiality of the host and interactions of the
debris with other satellites.

Our conclusions can be summarised as follows.  (1) The distributions
of (debris of) satellite galaxies in both real and integrals-of-motion
space are sensitive to the evolution (and the particulars) of their
host galaxy. This puts a caution on studies that investigate the shape
of the halo based on satellite streams obtained via simulations with
static DM halos.  (2) Even in a $\Lambda$CDM ``live'' halo, satellites
still appear to be coherent structures in the integrals-of-motion
space. However, the coherency is smeared significantly in contrast to
predictions from simulations using static and time-varying host
potentials, respectively.  Thus, earlier studies of the detection
ability of streams using a static DM potentials
\citep[e.g.][]{HdZ00,HMO01} and even time-varying potentials (Zhao~\ea 1999)
could overestimate the efficiency of the ``integrals-of-motion
approach''.  (3) In the integrals-of-motion space, energy changes most
significantly due to the mass growth of the host halo which is
inevitable in hierarchical structure formation scenarios.  However,
there are additional effects at work such as triaxiality of the host
and interactions of the stream with other satellites.  Hence, any
currently observed distribution of a satellite stream in the $E-L$
plane no longer reflects its original distribution.

\section{Acknowledgments} 
The simulations presented in this paper were carried out on the
Beowulf cluster at the Centre for Astrophysics~\& Supercomputing,
Swinburne University. The financial support of the Australian Research
Council is gratefully acknowledged.
We also appreciate useful discussion with Chris Brook.


\label{lastpage}

\end{document}